\documentclass[12pt]{article}
\usepackage{amsfonts}
\usepackage{mathrsfs}
\usepackage{amsfonts,amssymb,mathrsfs}
\usepackage{amsmath,amscd}
\usepackage[dvips]{graphicx}
\usepackage{color}
\usepackage[all]{xy}
\usepackage{graphicx}
\usepackage{mathptm,pslatex}
\usepackage{longtable}
\usepackage{supertabular}

\definecolor{red}{rgb}{1,0,0}
\definecolor{blue}{rgb}{0,0,1}
\definecolor{darkblue}{rgb}{0,0,0.4}

\begin{document}
\newtheorem{Def}{Definition}[section]
\newtheorem{exe}{Example}[section]
\newtheorem{prop}[Def]{Proposition}
\newtheorem{theo}[Def]{Theorem}
\newtheorem{lem}[Def]{Lemma}
\newtheorem{rem}{\noindent\mbox{Remark}}[section]
\newtheorem{coro}[Def]{Corollary}
 \newcommand{\Ker}{\rm Ker}
  \newcommand{\Soc}{\rm Soc}
 \newcommand{\lra}{\longrightarrow}
 \newcommand{\ra}{\rightarrow}
 \newcommand{\add}{{\rm add\, }}
\newcommand{\gd}{{\rm gl.dim\, }}
\newcommand{\End}{{\rm End\, }}
\newcommand{\overpr}{$\hfill\square$}
\newcommand{\rad}{{\rm rad\,}}
\newcommand{\soc}{{\rm soc\,}}
\renewcommand{\top}{{\rm top\,}}
\newcommand{\fdim}{{\rm fin.dim}\,}
\newcommand{\fidim}{{\rm fin.inj.dim}\,}
\newcommand{\gldim}{{\rm gl.dim}\,}
\newcommand{\cpx}[1]{#1^{\bullet}}
\newcommand{\D}[1]{{\mathscr D}(#1)}
\newcommand{\Dz}[1]{{\rm D}^+(#1)}
\newcommand{\Df}[1]{{\rm D}^-(#1)}
\newcommand{\Db}[1]{{\mathscr D}^b(#1)}
\newcommand{\C}[1]{{\mathscr C}(#1)}
\newcommand{\Cz}[1]{{\rm C}^+(#1)}
\newcommand{\Cf}[1]{{\rm C}^-(#1)}
\newcommand{\Cb}[1]{{\mathscr C}^b(#1)}
\newcommand{\K}[1]{{\mathscr K}(#1)}
\newcommand{\Kz}[1]{{\rm K}^+(#1)}
\newcommand{\Kf}[1]{{\rm K}^-(#1)}
\newcommand{\Kb}[1]{{\mathscr K}^b(#1)}
\newcommand{\modcat}[1]{#1\mbox{{\rm -mod}}}
\newcommand{\stmodcat}[1]{#1\mbox{{\rm -{\underline{mod}}}}}
\newcommand{\pmodcat}[1]{#1\mbox{{\rm -proj}}}
\newcommand{\imodcat}[1]{#1\mbox{{\rm -inj}}}
\newcommand{\opp}{^{\rm op}}
\newcommand{\otimesL}{\otimes^{\rm\bf L}}
\newcommand{\rHom}{{\rm\bf R}{\rm Hom}\,}
\newcommand{\pd}{{\rm pd}}
\newcommand{\Hom}{{\rm Hom \, }}
\newcommand{\Coker}{{\rm coker}\,\,}
\newcommand{\Ext}{{\rm Ext}}
\newcommand{\im}{{\rm im}}
\newcommand{\Cone}{{\rm Cone}}
\newcommand{\fini}{{finitistic}}
\newcommand{\proj}{{\rm  proj}}
\newcommand{\al}{{\alpha}}
\newcommand{\tr}{{ tr}}
\newcommand{\Tor}{{\rm  Tor}}
\newcommand{\ad}{{\rm  add}}
\newcommand{\StHom}{{\rm \underline{Hom} \, }}

\renewcommand{\theequation}{\arabic{section}.\arabic{equation}}
\pagenumbering{arabic}{\Large \bf
\begin{center}
The Jacobi sums over Galois rings of arbitrary characters and their applications in constructing asymptotically optimal codebooks
\end{center}}
\centerline{ {\sc Deng-Ming Xu$^1$ \qquad Chen Meng$^2$\qquad Gang Wang$^{3^*}$ \qquad Fang-Wei Fu$^4$ }}

\abstract{
 Codebooks with small maximum cross-correlation amplitudes are used to distinguish the signals from different users in CDMA communication systems. In this paper, we first study  the Jacobi sums over Galois rings of arbitrary  characteristics and completely determine their absolute values, which extends the work in \cite{feng1}, where the Jacobi sums over  Galois rings with
characteristics of a square of a prime number were discussed. Then, the deterministic construction of codebooks based on the Jacobi sums over Galois rings of arbitrary  characteristics is presented, which produces asymptotically optimal codebooks with respect to the Welch bound. In addition, the parameters of the codebooks provided in this paper are new.

\vskip -1.4cm
\renewcommand{\thefootnote}{\alph{footnote}}
\setcounter{footnote}{-1} \footnote{This work is supported by the National Natural Science Foundation of China under Grant 11701558 and the Doctoral Foundation of Tianjin Normal University under Grant 52XB2014.}
\renewcommand{\thefootnote}{\alph{footnote}}
\setcounter{footnote}{-1} \footnote{$^1$ Sino-European Institute of Aviation Engineering,
 Civil Aviation University of China, 300300, Tianjin,
People's Republic of  China. E-mail:
xudeng17@163.com.
}
\setcounter{footnote}{-1} \footnote{$^2$ College of Science,
 Civil Aviation University of China, 300300, Tianjin,
People's Republic of  China. E-mail: Mengchen$_{-}$246@163.com.
}
\setcounter{footnote}{-1} \footnote{$^3$ School of Mathematical Sciences, Tianjin Normal University, 300387, Tianjin,
People's Republic of  China. E-mail: gwang06080923@mail.nankai.edu.cn.
}
\setcounter{footnote}{-1} \footnote{$^4$ Chern Institute of Mathematics and LPMC, Nankai University, 300071, Tianjin,
People's Republic of China. E-mail: fwfu@nankai.edu.cn.
}
\setcounter{footnote}{-1} \footnote{$^*$Corresponding author, e-mail: gwang06080923@mail.nankai.edu.cn
}
\renewcommand{\thefootnote}{\alph{footnote}}
\setcounter{footnote}{-1} 
\renewcommand{\thefootnote}{\alph{footnote}}
\setcounter{footnote}{-1} \footnote{}

\section{Introduction}

The codebooks, whose maximal cross-correlation amplitudes are small, are applied to differentiate the signals from different participants in CDMA communication systems ({\cite{1}, \cite{2}}). Precisely, an $(N,K)$ codebook $\mathcal{C}$ is a set $\{ {c_0},{c_1}, \cdots ,{c_{N - 1}}\}$ of $N$ elements, where each ${c_i}\ (0 \le i \le N - 1)$ is a unit norm $1 \times K$ complex-valued vector over an alphabet $A$. The size of alphabet $A$ is called the alphabet size of the codebook $\mathcal{C}$.

 The maximal cross-correlation amplitude of an $(N,K)$ codebook $\mathcal{C}$, denoted by ${I_{\max }}(\mathcal{C})$, is defined by the following as a performance measure of the codebook,
  $${I_{\max }}(\mathcal{C}) = \mathop {\max }\limits_{0 \le i \ne j \le N - 1} \left| {{c_i}c_j^H} \right|
,$$ where $c_j^H$ is the conjugate transpose of the complex vector ${c_j}$. In practical applications, such $(N,K)$ codebooks $\mathcal{C}$ that $N$ is large and ${I_{\max }}(\mathcal{C})$ is small are desirable for a fixed $K$. However, the well-known Welch bound gives the lower bound for ${I_{\max }}(\mathcal{C})$.

\begin{lem}  {\cite{3}} \label{lemma 1.1} For any $(N,K)$ codebook $\mathcal{C}$ with $N > K$, $${I_{\max }}(\mathcal{C}) \ge {I_W} =\sqrt {\frac{{N - K}}{{(N - 1)K}}},$$ where the equality holds if and only if $$\left| {{c_i}c_j^H} \right| = \sqrt {\frac{{N - K}}{{(N - 1)K}}}$$ for any $0 \le i, j \le N - 1$, $i \ne j $.
\end{lem}

If the maximal cross-correlation amplitude of a codebook achieves the equality of the Welch bound, the codebook is called a maximum-Welch-bound-equality (MWBE) codebook. The MWBE codebooks have applications in many fields, such as CDMA communications ({\cite{1}}), space-time codes ({\cite{4}}) and compressed sensing ({\cite{5}}) and so on ({\cite{6}}, {\cite{7}}, {\cite{8}}). It is worthwhile to point out that constructing MWBE codebooks is very hard in general ({\cite{9}}). Only several classes of MWBE codebooks are provided till now in the previous literatures.

$\bullet$ The $(N,N)$ orthogonal codebooks and $(N,N - 1)$ MWBE codebooks from the discrete Fourier transform matrices or $m$-sequences ({\cite{10}});

$\bullet$ The $(N,K)$ MWBE codebooks from conference matrices, where $N = 2K = {2^{d + 1}}$ with a positive integer $d$ or $N = 2K = {p^d} + 1$ with a prime $p$ and a positive integer $d$ ({\cite{11})};

$\bullet$ The $(N,K)$ MWBE codebooks from $(N,K,\lambda )$ difference sets ({\cite{12}});

$\bullet$ The $(N,K)$ MWBE codebooks from $(2,k,\nu )$-Steiner systems ({\cite{13}});

$\bullet$ The $(N,K)$ MWBE codebooks from graph theory and finite geometries ({\cite{14}}).

Hence, many researchers attempt to construct asymptotically optimal codebooks, whose maximal cross-correlation amplitudes asymptotically achieve the corresponding Welch bound for sufficiently large $K$. The asymptotically optimal codebooks are constructed mainly based on combinatorial design ({\cite{15}}, {\cite{16}}), character sums ({\cite{17}}, {\cite{18}}), sequences ({\cite{Yu}}). For the sake of comparisons, Table 1 lists the parameters of the known asymptotically optimal codebooks with respect to the Welch bound in the existed literatures of recent years.

 \begin{table}
\caption{The parameters of codebooks asymptotically achieving Welch bound.}
\label{table}
\setlength{\tabcolsep}{9pt}
\begin{tabular}{|p{144pt}|p{73pt}|p{160pt}|p{50pt}|}
  \hline
  Parameters $(N,K)$ & ${I_{\max }}$ &  Constraints   & References   \\\hline
  $({p^n},K = \frac{{p - 1}}{{2p}}({p^n} + {p^{\frac{n}{2}}}) + 1)$  & $\frac{{(p + 1){p^{\frac{n}{2}}}}}{{2pK}}$ & $p$ is an odd prime   &{\cite{Hong}}   \\\hline
  $({q^2},\frac{{{{(q - 1)}^2}}}{2})$ & $\frac{{q + 1}}{{{{(q - 1)}^2}}}$ &  $q$ is an odd prime power    & {\cite{zhang}}           \\\hline
  $(q(q + 4),\frac{{(q + 3)(q + 1)}}{2})$ & $\frac{1}{{q + 1}}$ &    $q$ and $q+4$ are two prime powers &  {\cite{16}}  \\\hline
  $(q,\frac{{q - 1}}{2})$  & $\frac{{\sqrt q  + 1}}{{q - 1}}$ &  $q$ is a prime power   &   {\cite{16}} \\\hline
  $({q^l} + {q^{l - 1}} - 1,{q^{l - 1}})$  & $\frac{1}{{\sqrt {{q^{l - 1}}} }}$ &   $q$ is a prime power and $l > 2$   &   {\cite{15}} \\\hline
  $({(q - 1)^k} + {q^{k - 1}},{q^{k - 1}})$& $\frac{{\sqrt {{q^{k + 1}}} }}{{{{(q - 1)}^k} + {{( - 1)}^{k + 1}}}}$ & $q \ge 4$ is a prime power and $k > 2$  &    {\cite{17}} \\\hline
  $\begin{array}{l}
 ({(q - 1)^k} + K,K), \\
 K = \frac{{{{(q - 1)}^k} + {{( - 1)}^{k + 1}}}}{q} \\
 \end{array}$ & $\frac{{\sqrt {{q^{k - 1}}} }}{K}$ & $q$ is a prime power and $k > 2$  &   {\cite{17}} \\\hline
  $\begin{array}{l}
 ({({q^s} - 1)^n} + K,K), \\
 K = \frac{{{{({q^s} - 1)}^n} + {{( - 1)}^{n + 1}}}}{q} \\
 \end{array}$ & $\frac{{\sqrt {{q^{sn + 1}}} }}{{{{({q^s} - 1)}^n} + {{( - 1)}^{n + 1}}}}$ & $q$ is a prime power, $s > 1,n > 1$     &   {\cite{18}} \\\hline
  $({({q^s} - 1)^n} + {q^{sn - 1}},{q^{sn - 1}})$  & $\frac{{\sqrt {{q^{sn + 1}}} }}{{{{({q^s} - 1)}^n} + {{( - 1)}^{n + 1}}}}$ &$q$ is a prime power, $s > 1,n > 1$  &     {\cite{18}} \\\hline
  $({q^3} + {q^2},{q^2})$ & $\frac{1}{q}$ &  $q$ is a prime power  &    {\cite{Tian}} \\\hline
  $({q^3} + {q^2} - q,{q^2} - q)$  & $\frac{1}{{q - 1}}$ & $q$ is a prime power    &    {\cite{Tian}} \\\hline
  $({q^3} - q,{q^2} - q)$ & $\frac{1}{{q - 1}}$ &$q$ is a prime power     &   {\cite{Lu}}  \\\hline
  $({q^3} - 2q + 1,{(q - 1)^2})$  & $\frac{q}{{{{(q - 1)}^2}}}$ & $q$ is a prime power  &    {\cite{Lu}}  \\\hline
$(({p_{\min }} + 1){Q^2},{Q^2})$ & $\frac{1}{Q}$ & $Q > 1$ is an integer and ${p_{\min }}$ is the smallest prime factor of $Q$  &   {\cite{Qiu}} \\\hline

$(({p_{\min }} + 1){Q^2} - Q,Q(Q - 1))$ & $\frac{1}{{Q - 1}}$ &  $Q > 2$ is an integer and ${p_{\min }}$ is the smallest prime factor of $Q$   &   {\cite{Qiu}} \\\hline
$({N_1}{N_2},\frac{{{N_1}{N_2} - 1}}{2})$ & $\frac{{\sqrt {({N_1} + 1)({N_2} + 1)} }}{{{N_1}{N_2} - 1}}$ & ${N_1} \equiv 3\bmod 4$ and ${N_2} \equiv 3\bmod 4$   &   {\cite{Hu}} \\\hline
$({N_1} \cdots {N_l},\frac{{{N_1} \cdots {N_l} - 1}}{2})$ & $\frac{{\sqrt {({N_1} + 1) \cdots ({N_l} + 1)} }}{{{N_1} \cdots {N_l} - 1}}$ &${N_i} \equiv 3\bmod 4
$ for any $l \ge 1$ &   {\cite{Hu}} \\\hline
$\begin{array}{l}
 (2K + 1,K), \\
 K = \frac{{{{({2^{{s_1}}} - 1)}^n}{{({2^{{s_2}}} - 1)}^n} - 1}}{2} \\
 \end{array}$ & $\frac{{{2^{\frac{{{s_1}n + {s_2}n}}{2}}}}}{{{{({2^{{s_1}}} - 1)}^n}{{({2^{{s_2}}} - 1)}^n} - 1}}$ & $n \ge 1,{s_1},{s_2} > 1$  &   {\cite{Luo}} \\\hline
$\begin{array}{l}
 (2K + {( - 1)^{mn}},K), \\
 K = \frac{{{{({2^{{s_1}}} - 1)}^n} \cdots {{({2^{{s_m}}} - {{( - 1)}^{mn}})}^n} - 1}}{2} \\
 \end{array}$ & $\frac{{{2^{\frac{{{s_1}n + {s_2}n +  \cdots  + {s_m}n}}{2}}}}}{{2K}}$ & $n \ge 1,l > 1$ and ${s_i} > 1$ for any $1 \le i \le m$ &   {\cite{Luo}} \\\hline
$(k{p^2} + {p^2},{p^2})$ & $\frac{1}{p}$ & $p$ is a prime and $k|(p + 1)$  &   {\cite{Cao}} \\\hline
$({p^n} - 1,\frac{{{p^n} - 1}}{2})$ & $\frac{{\sqrt {{p^n}}  + 1}}{{{p^n} - 1}}$ & $p$ is an odd prime &   {\cite{Yu}} \\\hline
\end{tabular}
\label{tab1}
\end{table}

 Galois rings  play an  important  role in coding theory, combinatorics and cryptography  theory.
In \cite{ham},  sevral optimal nonliear binary codes were found  to be the  images of the Gray mapping of certain linear codes over $\mathbb{Z}_4.$  Since then, exponential sums over Galois rings have been widely used  in constructing error-correcting codes (\cite{car}, \cite {fengtao}), combinatorial designs (\cite{hou}, \cite{xi}) and mutually unbiased bases in quantum information theory (\cite{xu2}). The Jacobi sum is one of the most important  exponential sums over Galois rings. In \cite{feng1}, the authors presented explicit description of Jacobi sums  of two multiplicative characters over the Galois ring $GR(p^2,p^{2s})$.

 In this paper, we first study  the Jacobi sums over Galois rings of arbitrary  characteristics and completely determine their absolute values, which extends the work in \cite{feng1}. Then, the deterministic construction of codebooks based on the Jacobi sums over Galois rings of arbitrary of characteristics is presented, which produces asymptotically optimal codebooks with respect to the Welch bound. In addition, the parameters of the codebooks provided in this paper are new.

 The paper is organized as follows. We first give a brief introduction about Galois rings in Section 2 and then study the Jacobi sums of arbitrary  characteristics and their absolute values in Section 3. In Section 4, we present the construction of asymptotically optimal codebooks with the Jacobi sums over Galois rings. Finally, we give the  conclusion of the paper in Section 5.
\section{Preliminaries}

First, we collect some definitions, notations and basic results about Galois rings given in \cite{wan}. Let $p$ be a prime and $n\in\mathbb{N}$ such that $n\geq 2$.
Let $h(x)$ be a monic basic primitive polynomial of degree $s>0$ in $\mathbb{Z}_{p^n}[x]$ such that $h(x)|(x^{p^s-1}-1)$ in  $\mathbb{Z}_{p^n}[x]$. Then the Galois ring
$GR(p^n,p^{ns})$ of
characteristic $p^n$ is defined by the quotient ring $\mathbb{Z}_{p^n}[x]/(h(x))$. In this way, there is a root
$\xi$ of $h(x)$ with order $p^s-1$. Let $T=\{0,1,\xi^1,\xi^2,\cdots,\xi^{p^s-2}\}$ and $T^*=T\backslash\{0\}$.
Then $$GR(p^n,p^{ns})=\{a_0+a_1\xi+a_2\xi^2+\cdots+ a_{s-1}\xi^{s-1}\mid a_0,a_1,\cdots, a_{s-1}\in \mathbb{Z}_{p^n}\}.$$
Moreover, each $c\in GR(p^n,p^{ns})$ can be uniquely written as
$$c=c_0+pc_1+p^2c_2+\cdots+ p^{n-1}c_{{n-1}}$$
with $c_0,c_1,\cdots,c_{n-1}\in T$.
It is easy to see that $c$ is a unit in $GR(p^n,p^{ns})$ if and only if $c_0\neq 0$.

For abbreviation, set $q=p^s$, $R=GR(p^n,p^{ns}), M=pR$ and denote $R^*$ the set of units in $R$. Then  $R^*=T^*\times (1+M)$,
$|R|=q^n$,  $|R^*|=q^n-q^{n-1}$ and $|p^kR|=q^{n-k}$ for each $0\leq k\leq n$.
Let $1\leq k\leq n$. Set
$R_k=\{c_0+pc_1+p^2c_2+\cdots +p^{k-1}c_{{k-1}}\mid c_0,c_1,\cdots,c_{k-1}\in T\}$ and
$R_k^*=\{c_0+pc_1+p^2c_2+\cdots+ p^{k-1}c_{{k-1}}\mid c_0\in T^*,c_1,\cdots,c_{k-1}\in T\}$.
Then  each $c\in R^*$ can be uniquely written as $c=a+p^{k}b$ with $a\in R_k^*, b\in R_{n-k}.$

Define $\phi: R\rightarrow R$ as
$$a_0+a_1\xi+a_2\xi^2+\cdots+ a_{s-1}\xi^{s-1}\mapsto a_0+a_1\xi^p+a_2\xi^{2p}+\cdots+ a_{s-1}\xi^{(s-1)p}$$
for each $(a_0,a_1,a_2,\cdots, a_{s-1})\in (\mathbb{Z}_{p^n})^s$. Then $\phi$ is an automorphism of $R$ satisfying
$\phi(x)=x$ for each $x\in \mathbb{Z}_{p^n}.$
The generalized trace map ${\rm tr_n}$ from $R$ to $\mathbb{Z}_{p^n}$ is defined as
$x\mapsto x +\phi(x)+\cdots +\phi^{s-1}(x)$. The canonical additive character $\lambda $ on $R$ is defined as
$x\mapsto e^{\frac{2\pi i}{p^n}{\rm tr}_n(x)}$. For each $b\in R$, define $\lambda_b: R\mapsto \mathbb{C}^*$ as
$x\mapsto \lambda(bx)$. Then $\lambda_b$ is an additive character and $\{\lambda_b\mid b\in R\}$ contains all the additive characters
on $R$, we will denote it by $\widehat{R}$.

Let
$1\leq k\leq n-1$. Let $\tau_{n-k}$ be the map from $\mathbb{Z}_{p^n}$ to $\mathbb{Z}_{p^{n-k}}$ defined by $a\mapsto a\  ({\rm mod}\ p^{n-k}).$ Then it can
be extended to a map from $\mathbb{Z}_{p^n}[x]$ to $\mathbb{Z}_{p^{n-k}}[x]$. Let $f(x)\equiv h(x)\  ({\rm mod}\ p^{n-k})$.
Then $f(x)$ is also a monic primitive polynomial of degree $s$ dividing $x^{p^s-1}-1 $ in $\mathbb{Z}_{p^{n-k}}[x]$.
Then $\tau_{n-k}$  induces a map
$$\tau_{n-k}: GR(p^n,p^{ns})= \mathbb{Z}_{p^n}[x]/(h(x))\rightarrow  GR(p^{n-k},p^{(n-k)s})= \mathbb{Z}_{p^{n-k}}[x]/(f(x)).$$
Let ${\rm tr}_{n-k}$ be the generalized trace map from   $GR(p^{n-k},p^{(n-k)s})$ to $\mathbb{Z}_{p^{n-k}}$.
It is easy to check  that $\tau_{n-k}({\rm tr}_n(x))={\rm tr}_{n-k}(\tau_{n-k}(x))$ for each $x\in R$.

A multiplicative character on $R$ is a homomorphism  of groups from $R^*$ to $\mathbb{C}^*$.  The set of all multiplicative characters on $R$  will be denoted by $\widehat{R^*} $. We have the isomorphism of groups $${\tau _1}:(1 + {p^{n - 1}}R, \cdot ) \simeq ({\mathbb{F}_q}, + ),\ \ \ 1 + {p^{n - 1}}x \mapsto {\tau _1}(x)$$for $x \in T$. For any $a \in {\mathbb{F}_q}$, define ${\varphi _a}$ as ${\varphi _a}(1 + {p^{n - 1}}x) = {e^{\frac{{2\pi i}}{p}\mathrm{tr}(a{\tau _1}(x))}}$ for $x \in T$, where $\mathrm{tr}$ is the trace map from ${\mathbb{F}_q}$ to ${\mathbb{F}_p}$. Then, ${\varphi _a} \in \widehat{ 1 + {p^{n - 1}}R}$ and $\widehat{1 + {p^{n - 1}}R} = \left\{ {{\varphi _a}|a \in {\mathbb{F}_q}} \right\}$. We know that each multiplicative character $\chi  \in \widehat{ {R^ * }}$ can be written uniquely as $\chi  = \psi {\varphi _a}$, where $\psi  \in \widehat{ R_{n - 1,s}^ *} $ and $a \in {\mathbb{F}_q}$. Let $\chi\in \widehat{R^*}$.
Let  $1\leq k\leq n$. We say that $\chi $ is $k$-trivial if $\chi$  is trivial on $1+p^kR$ but non-trivial on $1+p^{k-1}R$.
If $\chi$ is trivial on $R$, we say that $\chi$ is $0$-trivial.
 Moreover, we say that $\chi$ is primitive if $\chi $ is $n$-trivial (compare to \cite{jun}).

 In the rest of the paper, we always denote $R$ the Galois ring $GR(p^n,p^{ns})$  and  $\chi_0$   the trivial multiplicative of $R$.

Let $\chi\in \widehat{R^*}$ and $\lambda\in \widehat{R}$. The Gauss sum of $\chi, \lambda$ is defined by
$$G(\chi,\lambda)=\displaystyle\sum_{x\in R^*}\chi(x)\lambda(x).$$

The following lemma can be checked directly by definition.
\begin{lem}\label{lem31}  Let $\chi\in \widehat{R^*}$ and $\lambda$ be the canonical additive character on $R$.  Let $a\in R$.
Then
$$G(\chi,\lambda_a)=
   \left\{\begin{array}{ll}
            q^n-q^{n-1}, & \chi=\chi_0, a=0;\\
            -q^{n-1},& \chi=\chi_0, a\neq 0, a\in p^{n-1}R;\\
            0,& \chi=\chi_0, a\notin p^{n-1}R;\\
            0,& \chi\neq \chi_0, a=0.
          \end{array}
   \right.$$
\end{lem}
The the following result can be deduced from Propositions 3.1 and 4.3 in  $\cite{lan}.$
\begin{lem} {\cite{lan}} \label{theo31} Let $\chi\in \widehat{R^*}$ be non-trivial and  $\lambda$ be the canonical additive character on $R$. Then we have the following:
 \begin{itemize}
   \item [$(1)$]  Let $a\in R^*$.  Then \qquad
   $|G(\chi,\lambda_a)|=
   \left\{\begin{array}{ll}
            q^{\frac{n}{2}}, & \chi\  is \ primitive;  \\
            0,& else.
          \end{array}
   \right.$
  \item [$(2)$] Let $a=p^{k}u$ with $u\in R_{n-k}^* $  and $1\leq k\leq n-1$. Then
  $$|G(\chi,\lambda_a)|=
   \left\{\begin{array}{ll}
            q^{\frac{n+k}{2}}, & \chi\  is\  (n-k)\mbox{-trivial}; \\
            0,& else.
          \end{array}
   \right.$$

 \end{itemize}
\end{lem}

\section{ The Jacobi sum over $GR(p^n,p^{ns})$}
Let $m\geq 2$.  Let  $\chi_1,\chi_2,\cdots,\chi_m $ be  $m$  multiplicative characters of $R$. Let $a\in R$.
The Jacobi sum of $\chi_1,\chi_2,\cdots,\chi_m $ is defined by
$$J_a(\chi_1,\chi_2,\cdots,\chi_m)=\displaystyle
 \sum_{\begin{smallmatrix}(x_1,\cdots,x_m)\in (R^*)^m\\
x_1+\cdots+x_m=a
\end{smallmatrix}}\chi_1(x_1)\chi_2(x_2)\cdots\chi_m(x_m).$$
Denote $J(\chi_1,\chi_2,\cdots,\chi_m)=J_1(\chi_1,\chi_2,\cdots,\chi_m)$.
As usual, we have
$$J_a(\chi_1,\chi_2,\cdots,\chi_m)
=\left\{\begin{array}{ll}
\chi_1\chi_2\cdots\chi_m(a)J(\chi_1,\chi_2,\cdots,\chi_m),& a\in R^*;\\
\chi_1\chi_2\cdots\chi_m(t)J_{p^k}(\chi_1,\chi_2,\cdots,\chi_m),& a=p^kt \ (1\leq k\leq n-1, t\in R_{n-k}^*).
\end{array}
\right.
$$
Consequently, it suffices to determine $J_a(\chi_1,\chi_2,\cdots,\chi_m)$ for $a\in \{0,1,p,\cdots, p^{n-1}\}$.

\subsection{Trivail cases}
The following lemma helps us determine $J_a(\chi_1,\chi_2,\cdots,\chi_m)$ for some trivial cases, the proof is given in \cite[Lemma 2.1]{D. Xu}.
\begin{lem} \label{lem41}Let $m\geq 2$ and $a\in R$. The number $S(n,r)^*$ of solutions $(c_1,c_2,\cdots,c_m)\in(R^*)^m$ of the equation
$$x_1+x_2+\cdots+x_m=a$$ is given by
$$S(n,r)^*
=\left\{\begin{array}{ll}
 q^{nm-m-n}\big((q-1)^m+(-1)^m(q-1)\big), &a\in M;\\
 q^{nm-m-n}\big((q-1)^m+(-1)^{m+1}\big),&a\notin M.
\end{array}
\right.
$$
\end{lem}

As a direct consequence of the previous lemma, we have

\begin{lem}\label{lem42} Let $m\geq 2.$
Let  $\chi_1,\chi_2,\cdots,\chi_m $ be  $m$ multiplicative characters of $R$.
If $\chi_1=\chi_2=\cdots=\chi_m=\chi_0 $, then
$$J_a(\chi_1,\chi_2,\cdots,\chi_m)=\left\{\begin{array}{ll}
 q^{nm-m-n}\big((q-1)^m+(-1)^m(q-1)\big),  &\mbox{if }\ a\in M;\\
 q^{nm-m-n}\big((q-1)^m+(-1)^{m+1}\big), & \mbox{if }\ a\notin M.
\end{array}
\right.$$
\end{lem}

Similar to the proof of \cite[Theorem 5.20]{nir}, one can prove the following result by Lemma \ref{lem41}.
\begin{lem}\label{lem43} Let $m\geq 3$.
Let  $\chi_1,\chi_2,\cdots,\chi_m \in \widehat{R^*}$ such that $\chi_m$ is non-trivial.
Then we have
$$J_0(\chi_1,\chi_2,\cdots,\chi_m)=\left\{\begin{array}{ll}
 0,  &\mbox{if }\chi_1\chi_2\cdots\chi_m \mbox{\ is \ non-trivial};\\
 \chi_m(-1)(q^n-q^{n-1})J(\chi_1,\chi_2,\cdots,\chi_{m-1}), &\mbox{if }\chi_1\chi_2\cdots\chi_m \mbox{\ is \ trivial}.
\end{array}
\right.$$
\end{lem}

Let  $\chi\in\widehat{R^*}$ and $S\subseteq R^*.$ If $\chi(s)=1$ for all $s\in S,$  then we write $\chi(S)=1$. Otherwise, we  write
$\chi(S)\neq 1$.

The proof of the following lemma  is similar to the work in \cite{feng1}, we omit it here.
\begin{lem} \label{lem44}
Let  $\chi_1,\chi_2\in\widehat{R^*}$. Then we have the following.

$(1)$
\begin{equation*}
  J(\chi_{1},\chi_{2})=
  \begin{cases}
  q^{n-1}(q-2),& if\quad \chi_1=\chi_2=\chi_0;\\
  0,& if \quad \chi_1=\chi_0,\ \chi_2(1+M)\neq 1 \ {or} \ \chi_2=\chi_0, \ \chi_1(1+M)\neq 1;\\
  -q^{n-1},& if \quad \chi_1=\chi_0,\ \chi_2\neq \chi_0,\ \chi_2(1+M)=1 \ or \ \chi_2=\chi_0,\ \chi _1\neq \chi_0,\ \chi_1(1+M)=1.\\
  \end{cases}
\end{equation*}

$(2)$\quad  Let $1\leq k\leq {n-1}$.
\begin{equation*}
  J_{p^k}(\chi_{1},\chi_{2})=
  \begin{cases}
  q^{n-1}(q-1),& if\quad \chi_1=\chi_2=\chi_0;\\
  0,& if \quad \chi_1=\chi_0,\ \chi_2\neq \chi_0\ \mbox{ or}\  \chi_2=\chi_0,\ \chi_1\neq \chi_0.\\
  \end{cases}
\end{equation*}
\end{lem}
\subsection{Jacobi sums of two multiplicative characters}

\begin{lem} \label{lem441}
Let  $\chi_1,\chi_2\in\widehat{R^*}$ be non-trivial such that $\chi_1\chi_2=1$.  Then we have the following.

$(1)$
\begin{equation*}
  J(\chi_{1},\chi_{2})=
  \begin{cases}
 - \chi_2(-1)q^{n-1},& if\ \chi_2(1+pR)=1;\\
0,& if\  \chi_2(1+pR)\neq 1.
  \end{cases}
\end{equation*}

$(2)$ Let  $1\leq k\leq {n-1}$.
\begin{equation*}
  J_{p^k}(\chi_{1},\chi_{2})=
  \begin{cases}
  0,& if \quad \chi_2(1+p^{k+1}R)\neq 1;\\
  \chi_2(-1)(q^{n}-q^{n-1}),& if \quad \chi_2(1+p^{k}R)=1;\\
 -\chi_2(-1)q^{n-1},& if \quad \chi_2\  is\  k+1\mbox{-trivial}. \\
  \end{cases}
\end{equation*}
\end{lem}
\noindent{\bf Proof} \quad (1) Since $\chi_1\chi_2=1$, we have
$$\begin{array}{lll}
J(\chi_1,\chi_2)&=&
\displaystyle \sum_{x\in R^*\backslash1+pR}\chi_1(x)\chi_2(1-x)=\displaystyle \sum_{x\in R^*\backslash1+pR}\chi_1\chi_2(x)\chi_2(x^{-1}-1)
\\
&=&\chi_2(-1)\displaystyle \sum_{x\in R^*\backslash1+pR}\chi_2(1-x)=\chi_2(-1)\displaystyle \sum_{y\in R^*\backslash1+pR}\chi_2(y)
\\
&=&\chi_2(-1)(\displaystyle \sum_{x\in R^*}\chi_2(y)-\displaystyle \sum_{y\in 1+pR}\chi_2(y))=-\chi_2(-1)\displaystyle \sum_{y\in 1+pR}\chi_2(y),
\end{array}
$$
where the last equation follows from $\chi_2\neq\chi_0.$ If $\chi_2(1+pR)\neq 1$, then $\displaystyle\sum_{y\in 1+pR}\chi_2(y)=0$ and therefore $J(\chi_1,\chi_2)=0$.
If $\chi_2(1+pR)=1$, then $\displaystyle\sum_{y\in 1+pR}\chi_2(y)=q^{n-1}$ and therefore $J(\chi_1,\chi_2)=-\chi_2(-1)q^{n-1}.$

(2) Similar to the proof of (1), we get that
$$\begin{array}{lll}
J_{p^k}(\chi_1,\chi_2)&=&\chi_2(-1)\displaystyle \sum_{x\in R^*}\chi_2(1+p^kx)=\chi_2(-1)\big(\displaystyle \sum_{x\in R}\chi_2(1+p^kx)-\displaystyle \sum_{x\in pR}\chi_2(1+p^{k}x)\big)\\
&=& \chi_2(-1)q^k
\big(\displaystyle \sum_{x\in R_{n-k}}\chi_2(1+p^kx)-\displaystyle \sum_{x\in R_{n-k-1}}\chi_2(1+p^{k+1}x)\big)\\
&=& \left\{
\begin{array}{ll}
  0, & \chi_2(1+p^{k+1}R)\neq 1; \\
  \chi_2(-1)(q^n-q^{n-1}), & \chi_2(1+p^{k}R)=1; \\
  -\chi_2(-1)q^{n-1}, & \chi_2\  is\  k+1\mbox{-trivial}.
\end{array}\right.
\end{array}$$
This finishes the proof of the lemma.\qquad $\square$

\begin{lem} \label{lem45}
Let $\chi_1,\chi_2\in\widehat{R^*}$  such that $\chi_2$ is primitive.
\begin{itemize}
  \item [$(1)$] Suppose $\chi_1\chi_2(1+p^{n-1}R)\neq1.$
Then $J_{p^k}(\chi_1,\chi_2)=0$ for $1\leq k\leq n-1.$
  \item [$(2)$] Suppose $\chi_1\chi_2(1+p^{n-1}R)=1.$
Then $J(\chi_1,\chi_2)=0$.
\end{itemize}

\end{lem}

\noindent{\bf Proof} \quad (1)  \ By definition, we have
$$\begin{array}{lll}J_{p^k}(\chi_1,\chi_2)=
\displaystyle \sum_{x\in R^*}\chi_1(x)\chi_2(p^k-x)=\displaystyle \sum_{x\in R^*}\chi_1\chi_2(x)\chi_2(p^kx^{-1}-1)
=\chi_2(-1)\displaystyle \sum_{x\in R^*}\overline{\chi_1\chi_2}(x)\chi_2(1-p^kx)
\end{array}
$$
Recall that for any $x\in R^*$, there exist $y\in R^*_{n-1}$
and $z\in R_1$ such that $x=y(1+p^{n-1}z)$.
Then  $1-p^kx=1-p^ky$. As a result,
$J_{p^k}(\chi_1,\chi_2)
=\chi_2(-1)\displaystyle \sum_{y\in R_{n-1}^*}\overline{\chi_1\chi_2}(y)\chi_2(1-p^ky)\displaystyle \sum_{z\in R_{1}}\overline{\chi_1\chi_2}(1-p^{n-1}z)=0
,$ where the last equation follows  from $\chi_1\chi_2(1+p^{n-1}R)\neq 1.$

(2) For abbreviation, set $A=\{x\in R_{n-1}^*\mid x\not\equiv 1 \ (\mbox{mod}\ p)\}$. Similar to the proof of (1), we have
$$\begin{array}{lll}
J(\chi_1,\chi_2)&=&
\displaystyle \sum_{x\in R^*\backslash1+pR}\chi_1(x)\chi_2(1-x)=\displaystyle \sum_{x\in A,y\in R_1}\chi_1(x(1+p^{n-1}y))\chi_2(1-x(1+p^{n-1}y))
\\
&=&\displaystyle \sum_{x\in A}\chi_1(x)\chi_2(1-x)\displaystyle \sum_{y\in R_1}\chi_1(1+p^{n-1}y)\chi_2(1-p^{n-1}(1-x)^{-1}xy))
\end{array}
$$
Note that $(1+p^{n-1}y)^{-1}=1-p^{n-1}y$. Since $\chi_1\chi_2(1+p^{n-1}R)=1,$ we get that
$$\displaystyle \sum_{y\in R_1}\chi_1(1+p^{n-1}y)\chi_2(1-p^{n-1}(1-x)^{-1}xy)=\displaystyle \sum_{y\in R_1}\chi_2(1-p^{n-1}y(1+x(1-x)^{-1})).$$
It is easy to see that for any $x\in A$, $1+x(1-x)^{-1}\notin pR$.  Since  $\chi_2(1+p^{n-1}R)\neq1,$  we get that $J(\chi_1,\chi_2)=0.$
\qquad $\square$

The following lemma is inspired by \cite[Lemma 2.4]{jun}.

\begin{lem} \label{lem46}Let $\lambda$ be the canonical additive character of $R$.
Let  $\chi_1,\chi_2\in\widehat{R^*}$  such that $\chi_2$ is primitive.
Suppose that $\chi_1\chi_2$ is $t$-trivial for some $1\leq t\leq n-1$.  Let $1\leq k\leq n-1$.
\begin{itemize}
  \item [$(1)$] If $n\neq t+k$,
then $J_{p^k}(\chi_1,\chi_2)=0$.
  \item [$(2)$] If $n=t+k$,
then $J_{p^k}(\chi_1,\chi_2)=\dfrac{q^k G(\chi_1,\lambda)G(\chi_2,\lambda)}{G(\chi_1\chi_2,\lambda_{p^k})}$.
\end{itemize}
\end{lem}

\noindent{\bf Proof} \quad (1)  \ By  the proof of Lemma \ref{lem45}, we have
$J_{p^k}(\chi_1,\chi_2)
=\chi_2(-1)\displaystyle \sum_{x\in R^*}\overline{\chi_1\chi_2}(x)\chi_2(1-p^kx).$
\begin{itemize}
  \item [(i)]Suppose that $ t+k>n$.  We  write $$J_{p^k}(\chi_1,\chi_2)=\chi_2(-1)\displaystyle \sum_{x\in R_{n-k}^*, \ y\in R_k}\overline{\chi_1\chi_2}(x(1+p^{n-k}y))\chi_2(1-p^kx(1+p^{n-k}y)).$$ Then we have
$$\displaystyle \sum_{x\in R_{n-k}^*, \ y\in R_k}\overline{\chi_1\chi_2}(x(1+p^{n-k}y))\chi_2(1-p^kx(1+p^{n-k}y))=
\displaystyle \sum_{x\in R_{n-k}^*}\overline{\chi_1\chi_2}(x)\chi_2(1-p^kx)\displaystyle \sum_{ y\in R_k}\overline{\chi_1\chi_2}(1+p^{n-k}y)$$
It follows from $\chi_1\chi_2$ is $t$-trivial and $t>n-k$ that
$\displaystyle \sum_{ y\in R_k}\overline{\chi_1\chi_2}(1+p^{n-k}y)=0$. Thus $J_{p^k}(\chi_1,\chi_2)=0$.
  \item [(ii)] Suppose that $ t+k<n$. We  write $$J_{p^k}(\chi_1,\chi_2)=\chi_2(-1)\displaystyle \sum_{x\in R_{t}^*, \ y\in R_{n-t}}\overline{\chi_1\chi_2}(x(1+p^{t}y))\chi_2(1-p^kx(1+p^{t}y)).$$ Since $\chi_1\chi_2$ is $t$-trivial, we get that
      $$J_{p^k}(\chi_1,\chi_2)=\chi_2(-1)\displaystyle \sum_{x\in R_{t}^*}\overline{\chi_1\chi_2}(x)\chi_2(1-p^kx)\sum_{\ y\in R_{n-t}}\chi_2(1-p^{k+t}x(1-p^{k}x)^{-1}y).$$
 Since $\chi_2$ is primitive, we get from  $k+t<n$ that $\displaystyle\sum_{\ y\in R_{n-t}}\chi_2(1-p^{k+t}x(1-p^{k}x)^{-1}y)=0.$
  Consequently,\\ $J_{p^k}(\chi_1,\chi_2)=0$.
\end{itemize}

(2) Suppose $k+t=n$.
 Firstly, we have $G(\chi_1,\lambda)G(\chi_2,\lambda)=\displaystyle\sum_{a\in R}\lambda(a)J_a(\chi_1,\chi_2).$ Since $\chi_1\chi_2$ is not primitive, we get from Lemma \ref{lem45} that
 $J_a(\chi_1,\chi_2)=0$ for all $a\in R^*.$ Meanwhile, we get from Lemma \ref{lem43} that $J_0(\chi_1,\chi_2)=0$ because $\chi_1\chi_2\neq \chi_0$. It follows from (1) that $J_{p^iu}(\chi_1,\chi_2)=0$
 for all $u\in R^*$ and  each $1\leq i\leq n-1$ satisfying $i\neq n-t$.
It follows that $$G(\chi_1,\lambda)G(\chi_2,\lambda)=\displaystyle\sum_{x\in R_t^*}\lambda(p^{n-t}x)J_{p^{n-t}x}(\chi_1,\chi_2)=
J_{p^{n-t}}(\chi_1,\chi_2)\displaystyle\sum_{x\in R_t^*}\lambda_{p^{n-t}}(x)\chi_1\chi_2(x)
.$$ Since $\chi_1\chi_2$ is $t$-trivial, we get  that $G(\chi_1\chi_2,\lambda_{p^{k}})=q^{n-t}\displaystyle\sum_{x\in R_t^*}\lambda_{p^{n-t}}(x)\chi_1\chi_2(x)$. Meanwhile, it follows  from Lemma \ref{theo31} that $G(\chi_1\chi_2,\lambda_{p^{k}})\neq 0$.
Consequently, $J_{p^k}(\chi_1,\chi_2)=q^k\dfrac{G(\chi_1,\lambda)G(\chi_2,\lambda)}{G(\chi_1\chi_2,\lambda_{p^k})}$.
\qquad $\square$

\begin{lem}\label{lem47} Let $\lambda$ be the canonical additive character of $R$.
Let  $\chi_1,\chi_2\in\widehat{R^*}$  such that $\chi_1\chi_2$ is primitive.
Then $J(\chi_1,\chi_2)=\dfrac{G(\chi_1,\lambda)G(\chi_2,\lambda)}{G(\chi_1\chi_2,\lambda)}.$
\end{lem}

\noindent{\bf Proof} \quad
Firstly, we have $G(\chi_1,\lambda)G(\chi_2,\lambda)=\displaystyle\sum_{a\in R}\lambda(a)J_a(\chi_1,\chi_2).$
It follows from Lemma \ref{lem45} that $J_a(\chi_1,\chi_2)=0$ for all $a\in M$. Then we have
$$G(\chi_1,\lambda)G(\chi_2,\lambda)=\displaystyle\sum_{a\in R^*}\lambda(a)J_a(\chi_1,\chi_2)=J(\chi_1,\chi_2)
\displaystyle\sum_{a\in R^*}\lambda(a)\chi_1\chi_2(a)=J(\chi_1\chi_2)G(\chi_1\chi_2,\lambda).$$
It follows from Lemma \ref{theo31} that $G(\chi_1\chi_2,\lambda)\neq 0$. Consequently,
$J(\chi_1,\chi_2)=\dfrac{G(\chi_1,\lambda)G(\chi_2,\lambda)}{G(\chi_1\chi_2,\lambda)}$.
\qquad $\square$

\subsection{General cases}

Let
$1\leq k\leq n-1$. Denote $R_{n-k,s}$ the Galois ring $GR(p^{n-k},p^{(n-k)s})$ for $1\leq k\leq n-1$.   Let $\chi\in\widehat{R^*}$  be $n-k$-trivial. Then
$\chi(a(1+p^{n-k}b))=\chi(a)$ for all  $a\in R_{n-k}^*$ and $ b\in R_k$.
Define a map $\widetilde{\chi}$ as follows
$$\forall\  \widetilde{a}\in R_{n-k,s}^*, \quad\widetilde{\chi}(\widetilde{a})=\chi(a),\quad  \ \mathrm{where}\ \tau_{n-k}(a)=\widetilde{a}.$$
Then
$\widetilde{\chi}$  is a multiplicative character of $R_{n-k,s}$. The following lemma can be proved directly.

\begin{lem} \label{lem481}
Let $m\geq 2$, $a\in R$ and $1\leq k\leq n-1$. Let  $\chi_1,\chi_2,\cdots,\chi_m \in\widehat{R^*}$ be  $n-k$-trivial.
Then $$J_a(\chi_1,\chi_2,\cdots,\chi_m)=q^{mk}\widetilde{J}_{\tilde{a}}(\widetilde{\chi}_1,\widetilde{\chi}_2,\cdots,\widetilde{\chi}_m),$$
where $\widetilde{J}_{\tilde{a}}(\widetilde{\chi}_1,\widetilde{\chi}_2,\cdots,\widetilde{\chi}_m)$ is the Jacobi sum of
$\widetilde{\chi}_1,\widetilde{\chi}_2,\cdots,\widetilde{\chi}_m$  over  $R_{n-k,s}$.
\end{lem}

Based on Lemma \ref{lem481},  it suffices to study the case that at least one of these characters
$\chi_1,\chi_2,\cdots, \chi_m \in\widehat{R^*}$
is primitive.
\begin{lem} \label{lem48}
Let $m\geq 3$. Let  $\chi_1,\chi_2,\cdots\chi_m \in\widehat{R^*}$  such that $\chi_m$ is primitive.
\begin{itemize}
  \item [$(1)$] Suppose that $\chi_1\chi_2\cdots\chi_m(1+p^{n-1}R)\neq1.$
Then $J_{p^k}(\chi_1,\chi_2,\cdots,\chi_m)=0$ for $1\leq k\leq n-1.$
  \item [$(2)$] Suppose that $\chi_1\chi_2\cdots\chi_m(1+p^{n-1}R)=1$.
Then $J(\chi_1,\chi_2,\cdots,\chi_m)=0$.
\end{itemize}
\end{lem}

\noindent{\bf Proof} \quad (1)  \ Let $1\leq k\leq n-1.$ For abbreviation, we always  assume  that $(a_1,a_2,\cdots,a_{m-1})\in (R^{\ast})^{m-1}$. By definition, we have
$$\begin{array}{lll}J_{p^k}(\chi_1,\chi_2,\cdots,\chi_m)
&=&\displaystyle \sum_{a_m\in R^*}\bigg(\displaystyle \sum_{a_1+\cdots+a_{m-1}=p^k-a_m}\chi_1(a_1)\cdots\chi_{m-1}(a_{m-1})\bigg)\chi_m(a_m)
\\
&=&
J(\chi_1,\cdots,\chi_{m-1})\displaystyle \sum_{a_m\in R^*}\chi_1\cdots\chi_{m-1}(p^k-a_m)\chi_m(a_m)
\\
&=&
J(\chi_1,\chi_2,\cdots,\chi_{m-1})J_{p^k}(\chi_1\cdots\chi_{m-1}, \chi_m)
\end{array}
$$
Since $\chi_1\chi_2\cdots\chi_m(1+p^{n-1}R)\neq1,$ we get from Lemma \ref{lem45}
 that $J_{p^k}(\chi_1\cdots\chi_{m-1}, \chi_m)=0$. Then (1) follows.

(2) By definition, we have
$$\begin{array}{lll}J(\chi_1,\chi_2,\cdots,\chi_m)&=&\displaystyle \sum_{a_m\in R^*\backslash 1+M}\bigg(\displaystyle \sum_{a_1+\cdots+a_{m-1}=1-a_m}\chi_1(a_1)\cdots\chi_{m-1}(a_{m-1})\bigg)\chi_m(a_m)
\\
&+&\displaystyle \sum_{a_m\in 1+M}\bigg(\displaystyle \sum_{a_1+\cdots+a_{m-1}=1-a_m}\chi_1(a_1)\cdots\chi_{m-1}(a_{m-1})\bigg)\chi_m(a_m)
\end{array}$$
 Firstly, $\displaystyle \sum_{a_m\in R^*\backslash 1+M}\bigg(\displaystyle \sum_{a_1+\cdots+a_{m-1}=1-a_m}\chi_1(a_1)\cdots\chi_{m-1}(a_{m-1})\bigg)\chi_m(a_m)
 =J(\chi_1,\chi_2,\cdots,\chi_{m-1})J(\chi_1\cdots\chi_{m-1},\chi_m)=0$,   where that last equation  follows from  the assumption $\chi_1\chi_2\cdots\chi_m(1+p^{n-1}R)=1$ and  Lemma \ref{lem45}.
On the other hand,    write $M= \bigg(\displaystyle\bigcup_{k=1}^{n-1}p^kR_{n-k}^*\bigg)\cup\{0\}$.
\begin{itemize}
  \item  If $a_m=1$, then $\bigg(\displaystyle \sum_{a_1+\cdots+a_{m-1}=1-a_m}\chi_1(a_1)\cdots\chi_{m-1}(a_{m-1})\bigg)\chi_m(a_m)=J_0(\chi_1,\chi_2,\cdots,\chi_{m-1})$. By the assumption, we get that
      $\chi_1\chi_2\cdots\chi_{m-1}$ is non-trivial. It follows from Lemma \ref{lem43} that this term equals to zero.
 \item By the assumption, we get that $\chi_1\chi_2\cdots\chi_{m-1}(1+p^{n-1}R)\neq 1$. It follows from (1) that $$J_{p^k}(\chi_1,\chi_2,\cdots,\chi_{m-1})=0.$$ Suppose that $1\leq k\leq n-1$. Then  we have equations

\hskip 2cm  $\begin{array}{lll}
 \displaystyle \sum_{a_m\in 1+p^kR_{n-k}^*}\bigg(\displaystyle \sum_{a_1+\cdots+a_{m-1}=1-a_m}\chi_1(a_1)\cdots\chi_{m-1}(a_{m-1})\bigg)\chi_m(a_m)
 \\
 =\displaystyle \sum_{x\in R_{n-k}^*}\bigg(\displaystyle \sum_{a_1+\cdots+a_{m-1}=p^kx}\chi_1(a_1)\cdots\chi_{m-1}(a_{m-1})\bigg)\chi_m(1+p^kx)
\\
=J_{p^k}(\chi_1,\chi_2,\cdots,\chi_{m-1})\displaystyle \sum_{x\in R_{n-k}^*}\chi_1\cdots\chi_{m-1}(x)\chi_m(1+p^kx)
=0.
\end{array}
$
\end{itemize}
 It follows that $\displaystyle \sum_{a_m\in 1+M}\bigg(\displaystyle \sum_{a_1+\cdots+a_{m-1}=1-a_m}\chi_1(a_1)\cdots\chi_{m-1}(a_{m-1})\bigg)\chi_m(a_m)=0.$ Consequently, $$J(\chi_1,\chi_2,\cdots,\chi_m)=0.\qquad \square$$

\begin{theo}\label{theo41} Let $m\geq 3$. Let  $\chi_1,\chi_2,\cdots,\chi_m \in\widehat{R^*}$  such that $\chi_m$ is  primitive.
Let $\lambda$ be the canonical additive character on $R$.
Then we have the following.
\begin{itemize}
  \item [$(1)$] $J(\chi_1,\chi_2,\cdots,\chi_m)=
  \begin{cases}
\dfrac{G(\chi_1,\lambda)\cdots G(\chi_m,\lambda)}{G(\chi_1\cdots\chi_m,\lambda)},& if\  \chi_1\chi_2\cdots\chi_m \ is\  \ primitive;\\
0,& if \ \chi_1\chi_2\cdots\chi_m \ is\ not  \ primitive.
  \end{cases}$
  \item  [$(2)$]  Suppose that
  $ \chi_1\chi_2\cdots\chi_m$ is $t$-trivial for $0\leq t\leq n$. Let $1\leq k\leq n-1$.
  Then
$$J_{p^k}(\chi_1,\chi_2,\cdots,\chi_m)=
  \begin{cases}
\dfrac{q^{k}G(\chi_1,\lambda)\cdots G(\chi_m,\lambda)}{G(\chi_1\cdots\chi_m,\lambda_{p^k})},& if\ 1\leq t\leq n-1,k=n-t;\\
-\chi_m(-1)q^{n-1}\dfrac{G(\chi_1,\lambda)\cdots G(\chi_{m-1},\lambda)}{G(\chi_1\cdots\chi_{m-1},\lambda)},& if\   t=0,k=n-1;\\
0,&else.
  \end{cases}$$
\end{itemize}

\end{theo}

\noindent{\bf Proof} \quad (1)
The second case follows from Lemma \ref{lem48}. Suppose that  $\chi_1\chi_2\cdots\chi_m $ is   primitive.
 Similar to the proof of \cite[Theorem 5.21]{nir}, we have
$$G(\chi_1,\lambda)\cdots G(\chi_m,\lambda)=J(\chi_1,\chi_2,\cdots,\chi_m)G(\chi_1\cdots\chi_m,\lambda).$$
It follows from Lemma \ref{theo31} that $G(\chi_1\cdots\chi_m,\lambda)\neq 0.$ Then the result follows.

(2) Let $1\leq k\leq n-1$. By the proof of  Lemma \ref{lem48} (1), we have
$$J_{p^k}(\chi_1,\chi_2,\cdots,\chi_m)=J(\chi_1,\chi_2,\cdots,\chi_{m-1})J_{p^k}(\chi_1\cdots\chi_{m-1}, \chi_m).$$
It follows from (1) and Lemma \ref{lem46} (2)  that the first case holds.
 The second case follows from Lemma \ref{lem441} (2).

It follows from Lemma \ref{lem46} (1) that $J_{p^k}(\chi_1,\chi_2,\cdots,\chi_m)=0$ when $1\leq t\leq n-1,k\neq n-t$. Since $\chi_m$ is primitive,  it follows from  Lemma \ref{lem441} (2) that $J_{p^k}(\chi_1,\chi_2,\cdots,\chi_m)=0$ when $t=0,k\neq n-1$.
  Finally, it follows from Lemma \ref{lem45} that $J_{p^k}(\chi_1,\chi_2,\cdots,\chi_m)=0$  when $t=n.$
\qquad $\square$

As a direct consequence of Lemma \ref{theo31} and Theorem \ref{theo41}, we get the following corollary.
\begin{coro}\label{coroo41} Let $m\geq 3$. Let  $\chi_1,\chi_2,\cdots,\chi_m \in\widehat{R^*}$  such that $\chi_m$ is  primitive.
Then we have the following.
\begin{itemize}
  \item [$(1)$] $|J(\chi_1,\chi_2,\cdots,\chi_m)|=
    \begin{cases}
q^{\frac{(m-1)n}{2}},& if\ each \ of \   \chi_1,\  \chi_2,\ \cdots,\ \chi_m,\    \ \chi_1\chi_2\cdots\chi_m \ is \  \ primitive;\\
0,& else.
  \end{cases}$
  \item  [$(2)$]  Suppose that
  $ \chi_1\chi_2\cdots\chi_m$ is $t$-trivial for $0\leq t\leq n$. Let $1\leq k\leq n-1$.
  Then
$$|J_{p^k}(\chi_1,\cdots,\chi_m)|=
  \begin{cases}
q^{\frac{(m-1)n+k}{2}},&
if\  1\leq t\leq n-1,k=n-t, \ and \ each \ of \   \chi_1,\  \chi_2,\ \cdots,\ \chi_m\ is \ primitive;\\
 q^{\frac{nm}{2}-1},&
 if\   t=0,k=n-1,  \ and
\  each\ of\  \chi_1,\  \cdots,\ \chi_{m-1},
   \ \chi_1\cdots\chi_{m-1} \ is \  \ primitive;
  \\
0,&else.
  \end{cases}$$
\end{itemize}

\end{coro}

\begin{rem} By Lemma \ref{lem481}, Corollary \ref{coroo41} and the results in Section 3.2, with the help of the absolute values of Jacobi sums over
 finite fields (see \cite[$\S$ 5.3]{nir}), we can completely determined
the absolute values of Jacobi sums over   any Galois ring.
\end{rem}

\section{The construction of asymptotically optimal codebooks}

 We extend the definition of a multiplicative character $\chi$ of $R$ by setting $\chi (M) = 1$ if $\chi$ is trivial and $\chi (M) = 0$ if $\chi$ is not trivial. Let $a \in R$ and $m \ge 2$. For a positive integer $k$ such that $1 \le k \le m - 1$, set $$S = \left\{ {({x_1}, \cdots ,{x_k},{x_{k + 1}}, \cdots ,{x_m}) \in {{({R^ * })}^k} \times {R^{m - k}}\mid{x_1} +  \cdots  + {x_k} + {x_{k + 1}} +  \cdots  + {x_m} = a} \right\}.$$The following lemma gives the cardinality of $S$.

\begin{lem} Let $a \in R$ and $m \ge 2$. For a positive integer $k$ such that $1 \le k \le m - 1$, denote $$S = \left\{ {({x_1}, \cdots ,{x_k},{x_{k + 1}}, \cdots ,{x_m}) \in {{({R^ * })}^k} \times {R^{m - k}}\mid{x_1} +  \cdots  + {x_k} + {x_{k + 1}} +  \cdots  + {x_m} = a} \right\}.$$Then $$\left| S \right| = \frac{{{{({q^n} - {q^{n - 1}})}^k} \cdot {{({q^n})}^{m - k}}}}{{{q^n}}}.$$
\end{lem}

\noindent{\bf Proof} \quad Let $\chi$ be the canonical additive character on $R$. Then $$\hspace {0.1cm} \left| S \right| = \frac{1}{{{q^n}}}\sum\limits_{y \in R} {\sum\limits_{\scriptstyle {x_1}, \cdots ,{x_k} \in {({R^ * })^k} \hfill \atop
  \scriptstyle {x_{k + 1}}, \cdots ,{x_m} \in {R^{m - k}} \hfill} {\chi (y({x_1} +  \cdots  + {x_k} + {x_{k + 1}} +  \cdots  + {x_m} - a))} }\hspace {7cm}$$
$$\hspace {0.6cm}   = \frac{1}{{{q^n}}}({({q^n} - {q^{n - 1}})^k} \cdot {({q^n})^{m - k}} + \sum\limits_{y \in R,y \ne 0} {\sum\limits_{\scriptstyle {x_1}, \cdots ,{x_k} \in {({R^ * })^k} \hfill \atop
  \scriptstyle {x_{k + 1}}, \cdots ,{x_m} \in {R^{m - k}} \hfill} {\chi ( - ay)} } \chi (y{x_1}) \cdots \chi (y{x_m}))\hspace {6cm}$$

$$\hspace {0.6cm} = \frac{1}{{{q^n}}}({({q^n} - {q^{n - 1}})^k} \cdot {({q^n})^{m - k}} +\sum\limits_{y \in R,y \ne 0} {\chi ( - ay)}  \cdot \sum\limits_{{x_1}, \cdots ,{x_k} \in {{({R^ * })}^k}} {\chi (y{x_1}) \cdots \chi (y{x_k})} \sum\limits_{{x_{k + 1}}, \cdots ,{x_m} \in {R^{m - k}}} {\chi (y{x_{k + 1}}) \cdots \chi (y{x_m})}\hspace {4cm}$$
$$\hspace {0.6cm} = \frac{{{{({q^n} - {q^{n - 1}})}^k} \cdot {{({q^n})}^{m - k}}}}{{{q^n}}},\hspace {15cm}$$where the last equation follows from $$\sum\limits_{{x_{k + 1}}, \cdots ,{x_m} \in {R^{m - k}}} {\chi (y{x_{k + 1}}) \cdots } \chi (y{x_m}) = \sum\limits_{{x_{k + 1}} \in R} {\chi (y{x_{k + 1}})}  \cdots \sum\limits_{{x_m} \in R} {\chi (y{x_m})}  = 0.\qquad \square$$

Let $a \in R$ and $m \ge 2$. For a positive integer $k$ such that $1 \le k \le m - 1$, we define a character sum by $${\tilde J_a}({\chi _1}, \cdots ,{\chi _k},{\chi _{k + 1}}, \cdots ,{\chi _m}) = \sum\limits_{({x_1}, \cdots ,{x_k},{x_{k + 1}}, \cdots ,{x_m}) \in S} {{\chi _1}({x_1}) \cdots {\chi _m}({x_m})},$$
where ${\chi _1}, \cdots ,{\chi _k},{\chi _{k + 1}}, \cdots ,{\chi _m}$ are $m$ multiplicative characters on $R$. The following lemma gives the relationship between the character sum ${\tilde J_a}({\chi _1}, \cdots ,{\chi _k},{\chi _{k + 1}}, \cdots ,{\chi _m})$ and Jacobi sum ${J_a}({\chi _1}, \cdots ,{\chi _m})$ over $R$.

\begin{lem} Let ${\tilde J_a}({\chi _1}, \cdots ,{\chi _k},{\chi _{k + 1}}, \cdots ,{\chi _m})$ be the character sum defined before.

(1) If all of ${\chi _{k + 1}}, \cdots ,{\chi _m}$ are nontrivial characters, then $${\tilde J_a}({\chi _1}, \cdots ,{\chi _k},{\chi _{k + 1}}, \cdots ,{\chi _m}) = {J_a}({\chi _1}, \cdots ,{\chi _m}).$$

(2) If ${\chi _{k + 1}}, \cdots ,{\chi _m}$ are trivial and ${\chi _1}, \cdots ,{\chi _k}$ are trivial, then $${\tilde J_a}({\chi _1}, \cdots ,{\chi _k},{\chi _{k + 1}}, \cdots ,{\chi _m}) = \frac{{{{({q^n} - {q^{n - 1}})}^k} \cdot {{({q^n})}^{m - k}}}}{{{q^n}}}.$$

(3) If ${\chi _{k + 1}}, \cdots ,{\chi _m}$ are trivial and some, but not all of ${\chi _1}, \cdots ,{\chi _k}$ are trivial, then $${\tilde J_a}({\chi _1}, \cdots ,{\chi _k},{\chi _{k + 1}}, \cdots ,{\chi _m}) = 0.$$

(4) If some, but not all of ${\chi _{k + 1}}, \cdots ,{\chi _m}$ are trivial, then $${\tilde J_a}({\chi _1}, \cdots ,{\chi _k},{\chi _{k + 1}}, \cdots ,{\chi _m}) = 0.$$
\end{lem}

\noindent{\bf Proof} \quad We only consider the case (4), i.e., some, but not all of ${\chi _{k + 1}}, \cdots ,{\chi _m}$ are trivial. For remaining cases the computations can be done easily.

Without loss of generality, assume that ${\chi _{k + 1}}, \cdots ,{\chi _{k + t}}$ are nontrivial and ${\chi _{k + t + 1}}, \cdots ,{\chi _m}$ are trivial for $1 \le t < m - k$. Then $${\tilde J_a}({\chi _1}, \cdots ,{\chi _k},{\chi _{k + 1}}, \cdots ,{\chi _m}) = (\sum\limits_{\scriptstyle {x_1}, \cdots ,{x_k} \in {({R^ * })^k} \hfill \atop
  {\scriptstyle {x_{k + 1}}, \cdots ,{x_m} \in {R^{m - k}} \hfill \atop
  \scriptstyle {x_1} +  \cdots  + {x_m} = a \hfill}} {{\chi _1}({x_1}) \cdots {\chi _{k + t}}({x_{k + t}})} ) \times \left| {M({x_1}, \cdots ,{x_{k + t}})} \right|,$$where $M({x_1}, \cdots ,{x_{k + t}}) = \left\{ {({x_{k + t + 1}}, \cdots ,{x_m}) \in {R^{m - k - t}}\mid{x_1} +  \cdots  + {x_{k + t}} +  \cdots  + {x_m} = a} \right\}$.

Then we get that $${{\tilde J}_a}({\chi _1}, \cdots ,{\chi _k},{\chi _{k + 1}}, \cdots ,{\chi _m}) = \sum\limits_{\scriptstyle {x_1}, \cdots ,{x_{k + t}} \in {({R^ * })^{k + t}} \hfill \atop
  \scriptstyle {x_1} +  \cdots  + {x_{k + t}} = a \hfill} {{\chi _1}({x_1}) \cdots {\chi _{k + t}}({x_{k + t}})}  \times {({q^n})^{m - k - t - 1}} + $$
$$\sum\limits_{\scriptstyle {x_1}, \cdots ,{x_{k + t}} \in {({R^ * })^{k + t}} \hfill \atop
  \scriptstyle {x_1} +  \cdots  + {x_{k + t}} \ne a \hfill} {{\chi _1}({x_1}) \cdots {\chi _{k + t}}({x_{k + t}})}  \times {({q^n})^{m - k - t - 1}}.$$Note that $$\sum\limits_{\scriptstyle {x_1}, \cdots ,{x_{k + t}} \in {({R^ * })^{k + t}} \hfill \atop
  \scriptstyle {x_1} +  \cdots  + {x_{k + t}} = a \hfill} {{\chi _1}({x_1}) \cdots {\chi _{k + t}}({x_{k + t}})}  + \sum\limits_{\scriptstyle {x_1}, \cdots ,{x_{k + t}} \in {({R^ * })^{k + t}} \hfill \atop
  \scriptstyle {x_1} +  \cdots  + {x_{k + t}} \ne a \hfill} {{\chi _1}({x_1}) \cdots {\chi _{k + t}}({x_{k + t}})}  = 0.$$

Hence, ${\tilde J_a}({\chi _1}, \cdots ,{\chi _k},{\chi _{k + 1}}, \cdots ,{\chi _m}) = 0.$\qquad $\square$

In the following, we present a construction of codebooks based on the character sum ${\tilde J_a}({\chi _1}, \cdots ,{\chi _k},{\chi _{k + 1}}, \cdots \\,{\chi _m})$ defined before and show that this class of codebooks is asymptotically optimal with respect to the Welch bound.

Let $a \in {R^ * }$, $m \ge 2$ and $k$ be a positive integer such that $1 \le k \le m - 1$. Let $K = \frac{{{{({q^n} - {q^{n - 1}})}^k} \cdot {{({q^n})}^{m - k}}}}{{{q^n}}}.$ Denote ${\mathcal{E}_K}$ the set formed by the standard basis of the $K$-dimensional Hilbert space:$$\begin{array}{*{20}{c}}
   {(1,0,0, \cdots ,0),}  \\
   {(0,1,0, \cdots ,0),}  \\
    \vdots   \\
   {(0,0, \cdots ,0,1).}  \\
\end{array}$$

For a fixed multiplicative character ${\psi _0} \in \widehat{ R_{n - 1,s}^ * }$, let ${\chi _1} = {\psi _0}{\varphi _{{a_1}}},{\chi _2} = {\psi _2}{\varphi _{{a_2}}}, \cdots ,{\chi _m}= {\psi _m}{\varphi _{{a_m}}}$ be $m$ multiplicative characters on $R$, where ${\psi _2}, \cdots ,{\psi _m} \in \widehat{ R_{n - 1,s}^ * },{a_1},{a_2}, \cdots ,{a_m} \in {\mathbb{F}_q}$. Note that $$S = \left\{ {({x_1}, \cdots ,{x_k},{x_{k + 1}}, \cdots ,{x_m}) \in {{({R^ * })}^k} \times {R^{m - k}}\mid{x_1} +  \cdots  + {x_k} + {x_{k + 1}} +  \cdots  + {x_m} = a} \right\}$$ and $\left| S \right| = K = \frac{{{{({q^n} - {q^{n - 1}})}^k} \cdot {{({q^n})}^{m - k}}}}{{{q^n}}}$ by Lemma 4.1. We define a codeword of length $K$ by $${c_{({\psi _0}{\varphi _{{a_1}}},{\psi _2}{\varphi _{{a_2}}}, \cdots ,{\psi _m}{\varphi _{{a_m}}})}} = \frac{1}{{\sqrt {\left| {S({c_{({\psi _0}{\varphi _{{a_1}}},{\psi _2}{\varphi _{{a_2}}}, \cdots ,{\psi _m}{\varphi _{{a_m}}})}})} \right|} }}{({\psi _0}{\varphi _{{a_1}}}({x_1}) \cdots {\psi _m}{\varphi _{{a_m}}}({x_m}))_{({x_1}, \cdots ,{x_m}) \in S}},$$where $S(c)$ denotes the support of a vector $c = ({c_1},{c_2}, \cdots ,{c_n})$ defined as $S(c) = \left\{ {1 \le i \le n:{c_i} \ne 0} \right\}$.

We know that the codeword ${c_{({\psi _0}{\varphi _{{a_1}}},{\psi _2}{\varphi _{{a_2}}}, \cdots ,{\psi _m}{\varphi _{{a_m}}})}}$ is a unit norm $1 \times K$ complex vector and then define a set by $$\mathcal{F} = \left\{ {{c_{({\psi _0}{\varphi _{{a_1}}},{\psi _2}{\varphi _{{a_2}}}, \cdots ,{\psi _m}{\varphi _{{a_m}}})}}\mid{\psi _2}, \cdots ,{\psi _m}\in \widehat{ R_{n - 1,s}^ * },{a_1},{a_2}, \cdots ,{a_m} \in {\mathbb{F}_q}} \right\}.$$Combining the set ${\mathcal{E}_K}$ and $\mathcal{F}$, we construct a codebook $\mathcal{C}$ as $$\hspace {6cm} \mathcal{C} = \mathcal{F} \cup {\mathcal{E}_K}.\hspace {8cm} (1) $$By definition, $\mathcal{C}$ has $N = q{({q^n} - {q^{n - 1}})^{m - 1}} + K$ codewords of length $K = \frac{{{{({q^n} - {q^{n - 1}})}^k} \cdot {{({q^n})}^{m - k}}}}{{{q^n}}}.$

\begin{theo} Let $a \in {R^ * }$, $m \ge 2$ and $k$ be a positive integer such that $1 \le k \le m - 1$. Then the set in Equation (1) is a $(q{({q^n} - {q^{n - 1}})^{m - 1}} + K,K)$ codebook with $${I_{\max }}(\mathcal{C}) = \frac{{{q^{\frac{{(m - 1)n}}{2}}}}}{{{q^{mn - m - n}}({{(q - 1)}^m} + {{( - 1)}^{m + 1}})}},$$where $K = \frac{{{{({q^n} - {q^{n - 1}})}^k} \cdot {{({q^n})}^{m - k}}}}{{{q^n}}}.$
\end{theo}

\noindent{\bf Proof} \quad By Lemma 4.1 and definition of $\mathcal{C}$, the size and length of codebook $\mathcal{C}$ are $q{({q^n} - {q^{n - 1}})^{m - 1}} + K$ and $K$, respectively, where $K = \frac{{{{({q^n} - {q^{n - 1}})}^k} \cdot {{({q^n})}^{m - k}}}}{{{q^n}}}.$ Let ${c_1},{c_2} \in \mathcal{C}$ be two distinct codewords. We calculate the correlation of ${c_1}$ and ${c_2}$ by distinguishing among the following cases.

(1) If ${c_1},{c_2} \in {\mathcal{E}_K}$, then $\left| {{c_1}c_2^H} \right| = 0$.

(2) If ${c_1} \in {\mathcal{E}_K},{c_2} \in \mathcal{F}$ or ${c_2} \in {\mathcal{E}_K},{c_1} \in \mathcal{F}$, then $$\left| {{c_1}c_2^H} \right| \in \left\{ {0,\frac{1}{{\left| {S({c_{({\psi _0}{\varphi _{{a_1}}},{\psi _2}{\varphi _{{a_2}}}, \cdots ,{\psi _m}{\varphi _{{a_m}}})}})} \right|}}} \right\}$$for some $({x_1},{x_2}, \cdots ,{x_m}) \in S$ and ${\psi _2}, \cdots ,{\psi _m} \in \widehat{ R_{n - 1,s}^ *}, {a_1},{a_2}, \cdots ,{a_m} \in {\mathbb{F}_q}$.

(3) If ${c_1},{c_2} \in \mathcal{F}$, we assume that $${c_1} = {c_{({\psi _0}{\varphi _{{a_1}}},{\psi _2}{\varphi _{{a_2}}}, \cdots ,{\psi _m}{\varphi _{{a_m}}})}} = \frac{1}{{\sqrt {\left| {S({c_1})} \right|} }}{({\psi _0}{\varphi _{{a_1}}}({x_1}) \cdots {\psi _m}{\varphi _{{a_m}}}({x_m}))_{({x_1}, \cdots ,{x_m}) \in S}}$$and$${c_2} = {c_{({\psi _0}{\varphi _{{b_1}}},{\phi _2}{\varphi _{{b_2}}}, \cdots ,{\phi _m}{\varphi _{{b_m}}})}} = \frac{1}{{\sqrt {\left| {S({c_2})} \right|} }}{({\psi _0}{\varphi _{{b_1}}}({x_1}) \cdots {\phi _m}{\varphi _{{b_m}}}({x_m}))_{({x_1}, \cdots ,{x_m}) \in S}},$$ where ${\psi _2}, \cdots ,{\psi _m},{\phi _2}, \cdots ,{\phi _m} \in \widehat{ R_{n - 1,s}^ * },{a_1},{a_2}, \cdots {a_m},{b_1},{b_2}, \cdots ,{b_m} \in {\mathbb{F}_q}$ and $({\psi _2}, \cdots ,{\psi _m},{a_1}, \cdots ,{a_m}) \ne ({\phi _2}, \cdots ,\\{\phi _m},{b_1}, \cdots ,{b_m})$. Then we have $$\left| {{c_1}c_2^H} \right| = \frac{{\left| {\sum\limits_{({x_1}, \cdots ,{x_m}) \in S} {{\varphi _{{a_1} - {b_1}}}({x_1}){\psi _2}\phi _2^{ - 1}{\varphi _{{a_2} - {b_2}}}({x_2}) \cdots {\psi _m}\phi _m^{ - 1}{\varphi _{{a_m} - {b_m}}}({x_m})} } \right|}}{{\sqrt {\left| {S({c_1})} \right|}  \cdot \sqrt {\left| {S({c_2})} \right|} }}$$

$$\hspace {2cm} = \frac{{\left| {{{\tilde J}_a}({\varphi _{{a_1} - {b_1}}},{\psi _2}\phi _2^{ - 1}{\varphi _{{a_2} - {b_2}}}, \cdots ,{\psi _m}\phi _m^{ - 1}{\varphi _{{a_m} - {b_m}}})} \right|}}{{\sqrt {\left| {S({c_1})} \right|}  \cdot \sqrt {\left| {S({c_2})} \right|} }}.\hspace {3.2cm}$$Since $({\psi _2}, \cdots ,{\psi _m},{a_1}, \cdots ,{a_m}) \ne ({\phi _2}, \cdots ,{\phi _m},{b_1}, \cdots ,{b_m})$, not all of ${\varphi _{{a_1} - {b_1}}},{\psi _2}\phi _2^{ - 1}{\varphi _{{a_2} - {b_2}}}, \cdots ,{\psi _m}\phi _m^{ - 1}{\varphi _{{a_m} - {b_m}}}$ are trivial. By Lemma 4.2, $$\left| {{c_1}c_2^H} \right| \in \left\{ {0,\frac{{\left| {{J_a}({\varphi _{{a_1} - {b_1}}},{\psi _2}\phi _2^{ - 1}{\varphi _{{a_2} - {b_2}}}, \cdots ,{\psi _m}\phi _m^{ - 1}{\varphi _{{a_m} - {b_m}}})} \right|}}{{\sqrt {\left| {S({c_1})} \right|}  \cdot \sqrt {\left| {S({c_2})} \right|} }}} \right\}.$$Furthermore, if ${a_1} \ne {b_1}$, then ${\varphi _{{a_1} - {b_1}}}$ is primitive and $\left| {{c_1}c_2^H} \right| \in \left\{ {0,\frac{{{q^{\frac{{(m - 1)n}}{2}}}}}{{\sqrt {\left| {S({c_1})} \right|}  \cdot \sqrt {\left| {S({c_2})} \right|} }}} \right\}$ by Corollary 3.12. If ${a_1} = {b_1}$, then ${\varphi _{{a_1} - {b_1}}} = {\varphi _0}$ is trivial and $\left| {{c_1}c_2^H} \right| = 0$ by Lemma \ref{lem481} and Corollary \ref{coroo41}.

Note that for any ${c_{({\psi _0}{\varphi _{{a_1}}},{\psi _2}{\varphi _{{a_2}}}, \cdots ,{\psi _m}{\varphi _{{a_m}}})}} \in \mathcal{F}$ where ${\psi _2}, \cdots ,{\psi _m} \in \widehat{ R_{n - 1,s}^ *} ,{a_1}, \cdots ,{a_m} \in {\mathbb{F}_q}$, $${q^{mn - m - n}}({(q - 1)^m} + {( - 1)^{m + 1}}) \le \left| {S({c_{({\psi _0}{\varphi _{{a_1}}},{\psi _2}{\varphi _{{a_2}}}, \cdots ,{\psi _m}{\varphi _{{a_m}}})}})} \right| \le \frac{{{{({q^n} - {q^{n - 1}})}^k} \cdot {{({q^n})}^{m - k}}}}{{{q^n}}}.$$Therefore, $\left| {{c_1}c_2^H} \right| \le \frac{{{q^{\frac{{(m - 1)n}}{2}}}}}{{{q^{mn - m - n}}({{(q - 1)}^m} + {{( - 1)}^{m + 1}})}}$. Moreover, if ${\psi _{k + 1}}{\varphi _{{a_{k + 1}}}}, \cdots ,{\psi _m}{\varphi _{{a_m}}},{\phi _{k + 1}}{\varphi _{{b_{k + 1}}}}, \cdots ,{\phi _m}{\varphi _{{b_m}}}$ are nontrivial and each of ${\varphi _{{a_1} - {b_1}}},{\psi _2}\phi _2^{ - 1}{\varphi _{{a_2} - {b_2}}}, \cdots ,{\psi _m}\phi _m^{ - 1}{\varphi _{{a_m} - {b_m}}},({\varphi _{{a_1} - {b_1}}})({\psi _2}\phi _2^{ - 1}{\varphi _{{a_2} - {b_2}}}) \cdots ({\psi _m}\phi _m^{ - 1}{\varphi _{{a_m} - {b_m}}})$ is primitive, the equality holds.

Summarizing the cases above, we obtain $${I_{\max }}(\mathcal{C}) = \frac{{{q^{\frac{{(m - 1)n}}{2}}}}}{{{q^{mn - m - n}}({{(q - 1)}^m} + {{( - 1)}^{m + 1}})}}.\qquad \square$$

\begin{theo} Let $a \in {R^ * }$, $m \ge 2$ and $k$ be a positive integer such that $1 \le k \le m - 1$. The codebook $\mathcal{C}$ in Theorem 4.3 is asymptotically optimal with respect to the Welch bound.
\end{theo}

\noindent{\bf Proof} \quad Recall that $N = q{({q^n} - {q^{n - 1}})^{m - 1}} + K$ and $K = \frac{{{{({q^n} - {q^{n - 1}})}^k} \cdot {{({q^n})}^{m - k}}}}{{{q^n}}}$. The corresponding Welch bound of $\mathcal{C}$ is $${I_W} = \sqrt {\frac{{N - K}}{{(N - 1)K}}}  = \sqrt {\frac{{{q^{k - m + 2}}{{(q - 1)}^{m - k - 1}}}}{{q{{({q^n} - {q^{n - 1}})}^{m - 1}} + {{(q - 1)}^k}{q^{mn - k - n}} - 1}}}.$$By Theorem 4.3, $$\mathop {\lim }\limits_{q \to \infty } \frac{{{I_W}}}{{{I_{\max }}(\mathcal{C})}} = \mathop {\lim }\limits_{q \to \infty } \sqrt {\frac{{{q^{mn - 3m - n + k + 2}}{{(q - 1)}^{m - k - 1}}{{({{(q - 1)}^m} + {{( - 1)}^{m + 1}})}^2}}}{{q{{({q^n} - {q^{n - 1}})}^{m - 1}} + {{(q - 1)}^k}{q^{mn - k - n}} - 1}}}  = 1,$$which implies that the codebook $\mathcal{C}$ is asymptotically optimal with respect to the Welch bound. \qquad $\square$

In Table 2, for $n = 2, m = 3, k=1$ and some $q$, we obtain the explicit parameter values of the codebook $\mathcal{C}$ presented in Theorem 4.3. Meanwhile, we also obtain the corresponding Welch bound ${I_W}$
and ratio of $\frac{{{I_{\max }}(\mathcal{C})}}{{{I_W}}}$. It can be seen that the codebooks $\mathcal{C}$ asymptotically achieve the Welch bound.

\begin{rem} When $a = 0$, the codebook $\mathcal{C}$ in (1) has the parameters $(N = q{({q^n} - {q^{n - 1}})^{m - 1}} + K,K)$ and $${I_{\max }}(\mathcal{C}) = \frac{{({q^n} - {q^{n - 1}}){q^{\frac{{(m - 2)n}}{2}}}}}{{{q^{mn - m - n}}({{(q - 1)}^m} + {{( - 1)}^m}(q - 1))}},$$where $K = \frac{{{{({q^n} - {q^{n - 1}})}^k} \cdot {{({q^n})}^{m - k}}}}{{{q^n}}}.$When $a \in M\backslash \left\{ 0 \right\}$, the codebook $\mathcal{C}$ in (1) has the parameters $(N = q{({q^n} - {q^{n - 1}})^{m - 1}} + K,K)$ and $${I_{\max }}(\mathcal{C}) = \frac{{\sqrt {{q^{2m + 2n - mn - 1}}} }}{{{{(q - 1)}^m} + {{( - 1)}^m}(q - 1)}},$$where $K = \frac{{{{({q^n} - {q^{n - 1}})}^k} \cdot {{({q^n})}^{m - k}}}}{{{q^n}}}.$ The codebook $\mathcal{C}$ in both of the two cases are not optimal or asymptotically optimal with respect to the Welch bound.
\end{rem}

 \begin{table}
\caption{The explicit parameter values of codebooks in Theorem 4.3.}
\label{table}
\setlength{\tabcolsep}{10pt}
\begin{tabular}{|c|c|c|c|c|c|}
  \hline
  $q$ & $N$ & $K$ & ${I_{\max }}(\mathcal{C})$ & ${I_W}$ & $\frac{{{I_{\max }}(\mathcal{C})}}{{{I_W}}}$\\\hline
  11 & 146410 & 13310 & 0.010989011 & 0.008264491 & 1.329665789 \\\hline
  19 & 2345778 & 123462 & 0.003257329 & 0.002770084 &1.175895515 \\\hline
  31 & 27705630 & 893730 & 0.0011481056 & 0.0010405827 & 1.1033294864 \\\hline
  53 & 410305012 & 7741604 & 0.0003769318 & 0.0003559986 & 1.0588013557 \\\hline
  81 & 3443737680 & 42515280 & 0.0001582028 &0.0001524158 & 1.0379686757 \\\hline
  121 & 25723065720 & 212587320 & 0.0000700231 & 0.0000683013 & 1.0252083187 \\\hline
  179 & 182739371218 & 1020890342& 0.00003173898 & 0.00003121001 & 1.01694861459 \\\hline
  256 & 1095216660480& 4278190080 & 0.00001543901 & 0.00001525879 & 1.01181084127 \\\hline
\end{tabular}
\label{tab1}
\end{table}

\section{Conclusion}

In this paper, we first study the  Jacobi sums over Galois rings of arbitrary  characteristics and determine their absolute values, which extends the  work given in \cite{feng1}. Then, the deterministic construction of codebooks based on the Jacobi sums over Galois rings of arbitrary  characteristics is presented, which produces asymptotically optimal codebooks with respect to the Welch bound. In addition, the $(q{({q^n} - {q^{n - 1}})^{m - 1}} + K,K)$ codebooks $\mathcal{C}$ with ${I_{\max }}(\mathcal{C}) = \frac{{{q^{\frac{{(m - 1)n}}{2}}}}}{{{q^{mn - m - n}}({{(q - 1)}^m} + {{( - 1)}^{m + 1}})}}$ provided in this paper are new, where $K = \frac{{{{({q^n} - {q^{n - 1}})}^k} \cdot {{({q^n})}^{m - k}}}}{{{q^n}}}$. We firmly believe that the Jacobi sums over Galois rings will have similar applications as the Jacobi sums over finite fields have.


{\footnotesize
\begin{thebibliography}{99}


\bibitem{1}J. Massey and T. Mittelholzer,$^{''}$Welch$^{\prime}$s bound and
sequence sets for code-division multiple-access systems,$^{''}$ in {\small{{\it Sequences II: Methods
in Communication, Security and Computer Science}}}, Springer, New York, pp. 63-78, 1999.

\bibitem{2}D. J. Love, R. W. Heath and T. Strohmer,$^{''}$Grassmannian beamforming
for multiple input multiple output wireless systems,$^{''}${\small{{\it IEEE Trans. Inf. Theory}}}, vol. 49, no. 10,
pp. 2735-2747, 2003.

\bibitem{3}L. Welch,$^{''}$Lower bounds on the maximum cross correlation
of signals,$^{''}${\small{{\it IEEE Trans. Inf. Theory}}}, vol. 20, no.
3, pp. 397-399, 1974.

\bibitem{4}V. Tarokh and I. M. Kim,$^{''}$Existence and construction of noncoherent
unitary space-time codes,$^{''}${\small{{\it IEEE Trans. Inf. Theory}}}, vol. 48, no. 12, pp. 3112-3117,
2002.

\bibitem{5}S. Li and G. Ge,$^{''}$Deterministic sensing matrices arising from
near orthogonal systems,$^{''}${\small{{\it IEEE Trans. Inf. Theory}}}, vol. 60, no. 4,
pp. 2291-2302, 2014.

\bibitem{6}J. Conway, R. Harding and N. Sloane,$^{''}$Packing lines,
planes, etc.: Packings in Grassmannian spaces,$^{''}${\small{{\it Exp.
Math.}}}, vol. 5, no. 2, pp. 139-159,
1996.

\bibitem{7}K. K. Mukkavilli, A. Sabharwal, E. Erkip et al.,$^{''}$On beamforming with finite rate feedback in multiple-antenna systems,$^{''}${\small{{\it IEEE Trans. Inf. Theory}}}, vol. 49, no. 10, pp. 2562-2579, 2003.

\bibitem{8}P. Delsarte, J. Goethals and J. Seidel,$^{''}$Spherical codes
and designs,$^{''}${\small{{\it
Geometriae Dedicate}}}, vol. 67, no. 3, pp. 363-388,
1997.

\bibitem{9}D. Sarwate,$^{''}$Meeting the Welch bound with equality,$^{''}$ in {\small{{\it Proc. SETA$^{\prime}$98}}}, London, U.K., pp. 79-102, 1999.

\bibitem{Hong}S. Hong, H. Park, T. Helleseth et al.,$^{''}$Near optimal partial Hadamard codebook construction using
binary sequences obtained from quadratic residue mapping,$^{''}${\small{{\it IEEE Trans. Inf. Theory}}}, vol. 60, no.
6, pp. 3698-3705, 2014.

\bibitem{zhang}A. Zhang and K. Feng,$^{''}$Two classes of codebooks nearly meeting the Welch bound,$^{''}${\small{{\it IEEE Trans. Inf. Theory}}}, vol. 58, no.
4, pp. 2507-2511, 2012.

\bibitem{Tian}L. Tian, Y. Li, T. Liu, et al.,$^{''}$Constructions of codebooks asymptotically achieving the Welch
bound with additive characters,$^{''}${\small{{\it IEEE Signal Pro. Let.}}}, vol. 26, no.
4, pp. 622-626, 2019.


\bibitem{Cao}G. Luo and X. Cao,$^{''}$New constructions of codebooks asymptotically achieving the Welch bound,$^{''}$ {\small{{\it IEEE Int. Symp. Inf. Theory}}}, Vail, CO, USA, pp. 2346-2350, 2018.




\bibitem{10}P. Xia, S. Zhou and G. Giannakis,$^{''}$Achieving the Welch
bound with difference sets,$^{''}${\small{{\it IEEE Trans. Inf. Theory}}}, vol. 51, no.
5, pp. 1900-1907, 2005.

\bibitem{11}T. Strohmer and R. Heath,$^{''}$Grassmannian frames with
applications to coding and communication,$^{''}${\small{{\it Appl. Comput.
Harmon. Anal.}}}, vol. 14, no. 3, pp. 257-275, 2003.

\bibitem{12}C. Ding,$^{''}$Complex codebooks from combinatorial designs,$^{''}${\small{{\it IEEE Trans. Inf. Theory}}}, vol. 52, no. 9, pp. 4229-4235, 2006.

\bibitem{13}M. Fickus, D. Mixon and J. Tremain,$^{''}$Steiner equiangular
tight frames,$^{''}${\small{{\it Linear Algebra Appl.}}}, vol. 436, no. 5, pp. 1014-1027, 2012.

\bibitem{14}M. Fickus, D. Mixon and J. Jasper,$^{''}$Equiangular tight
frames from hyperovals, $^{''}${\small{{\it IEEE Trans. Inf.
Theory}}}, vol. 62, no. 9, pp. 5225-5236, 2016.

\bibitem{15}Z. Zhou and X. Tang,$^{''}$New nearly optimal codebooks from
relative difference sets,$^{''}${\small{{\it Adv. Math. Commun.}}}, vol. 5, no. 3, pp. 521-527, 2011.

\bibitem{16}C. Li, Q. Yue and Y. Huang,$^{''}$Two families of nearly optimal codebooks,$^{''}${\small{{\it Des. Codes Cryptogr.}}}, vol. 75, no. 1, pp. 43-57, 2015.

\bibitem{17}Z. Heng, C. Ding and Q. Yue,$^{''}$New constructions of
asymptotically optimal codebooks with multiplicative
characters,$^{''}${\small{{\it IEEE Trans. Inf.
Theory}}}, vol. 63, no. 10, pp. 6179-6187, 2017.

\bibitem{18}G. Luo and X. Cao,$^{''}$Two constructions of asymptotically optimal codebooks via the hyper Eisenstein sum,$^{''}${\small{{\it IEEE Trans. Inf.
Theory}}}, vol. 64, no. 10, pp. 6498-6505, 2017.


\bibitem{Yu}N. Y. Yu,$^{''}$A construction of codebooks associated with binary sequences,$^{''}${\small{{\it IEEE Trans. Inf.
Theory}}}, vol. 58, no. 8, pp. 5522-5533, 2012.


\bibitem{D. Xu}D. Xu and C. Meng,$^{''}$Construction of asymptotically optimal codebooks over Galois rings,$^{''}$ {\small{{\it preprint}}}.


\bibitem{Luo}G. Luo and X. Cao,$^{''}$Two constructions of asymptotically optimal codebooks,$^{''}${\small{{\it Cryptogr. Commun.}}}, vol. 11, no. 2, pp. 825-838, 2019.


\bibitem{Lu}W. Lu, X. Wu, X. Cao, et al.,$^{''}$Six constructions of asymptotically optimal codebooks via the character sums,$^{''}${\small{{\it Des. Codes Cryptogr}}}, vol. 88, no. 2, pp. 1139-1158, 2020.


\bibitem{Qiu}Q. Wang and Y. Yan,$^{''}$Asymptotically optimal codebooks derived from generalised bent
functions,$^{''}${\small{{\it IEEE Access}}}, vol. 8, pp. 54905-54909, 2020.

\bibitem{Hu}H. Hu and J. Wu,$^{''}$New constructions of codebooks nearly meeting the
Welch bound with equality,$^{''}${\small{{\it IEEE Trans. Inf.
Theory}}}, vol. 60, no. 2, pp. 1348-1355, 2014.


 \bibitem{ham}A. R. Hammons, P. V. Kumar, A. R. Calderbank, et al., $^{''}$The $\mathbb{Z}_4$-linearity of Kerdock, Preparata, Goethals, and related codes$^{''}${\small{{\it IEEE Trans. Inf. Theory}}},vol. 40, no. 12, pp. 301-319, 1994.



\bibitem{car} C. Carlet, $^{''}$One-weight $\mathbb{Z}_4$-linear codes,$^{''}$in {\small{{\it  Proc. of an International Conference on Coding Theory Cryptography and Related Areas}}}, Guanajuato, Mexico, 57-72, 2000.


\bibitem{fengtao} T. Feng,$^{''}$ A new construction of perfect nonlinear functions using Galois rings,$^{''}$ \emph{ J. Combin. Designs,} vol. 17, no. 3, pp. 229-239, 2009.


 \bibitem{hou} X. Hou, K. Leung and Q. Xiang,$^{''}$ New partial difference sets in $\mathbb{Z}_{p^2}^t$ and a related problem
about Galois rings,$^{''}$ \emph{Finite Fields  Appl.}, vol. 7, no. 1, pp. 165-188, 2001.


 \bibitem{lan} P. Langevin and P. Sol\'e,$^{''}$ Gauss sums over quasi-Frobenius rings,$^{''}$ in {\small{{\it Proc. of the fifth international conference on finite fields and applications}}}, 329-340, 1999.



 \bibitem{feng1} J. Li, S. Zhu and K. Feng,$^{''}$ The Gauss sums and Jacobi sums over Galois ring $GR(p^2,r)$,$^{''}$\emph{ Sicence China}, vol. 56, no. 7, pp. 1457-1465, 2013.

 \bibitem{jun}J. Wang, $^{''}$On the Jacobi sums modulo $p^n$,$^{''}$ \emph{J. Number Theory,} vol. 39, no. 1, pp. 50-64, 1991.

 \bibitem{nir} R. Lidl and H. Niederreiter, \emph{Finite fields}, Cambridge University Press, 2nd edition, 1997.





 \bibitem{wan} Z. Wan,\emph{ Lectures Notes on Finite Fields and Galois Rings,} Word Scientific, Singapore, 2003.






 \bibitem{xi} Q. Xiang and  J. A. Davis, $^{''}$Constructions of low rank relative differrenecce sets in $2$-groups using Galois rings,$^{''}$ \emph{Finite Fields  Appl.} vol. 6, no. 2, pp. 130-145, 2000.



\bibitem{xu2} D. Xu,$^{''}$Construction of mutually unbiased maximally entangled bases in $\mathbb{C}^{2^s}\otimes\mathbb{C}^{2^s}$ by using Galois rings,$^{''}$ \emph{Quantum Inf. Process,} vol. 19, no. 6, 175, 2020.


\end{thebibliography}

}
\end{document}